\documentclass[aps,preprint,epsfig,rotate]{revtex4}
\begin{document}
\title{Hyperfine structure splitting of the ground states
       in the $pd\mu, pt\mu$ and $dt\mu$ ions}

 \author{Alexei M. Frolov}
 \email[E--mail address: ]{afrolov@uwo.ca}

\affiliation{Department of Chemistry\\
 University of Western Ontario, London, Ontario N6H 5B7, Canada}

\date{\today}

\begin{abstract}

The hyperfine structure splittings of the ground states of the $pd\mu, 
pt\mu$ and $dt\mu$ ions are determined with the use of highly accurate 
expectation values of the interparticle delta-functions obtained in 
recent computations. The corresponding hyperfine structure splittings, 
e.g., $\Delta_{12} =$ 1.3400149$\cdot 10^7$ $MHz$ and $\Delta_{23} =$ 
3.3518984$\cdot 10^7$ $MHz$ for the $pt\mu$ ion, can directly be measured 
in modern experiments.

\end{abstract}

\maketitle
\newpage

In this study we analyze the hyperfine structure and determine the hyperfine 
structure splitting of the bound $S(L = 0)$-states in the non-symmetric 
muonic molecular ions $pd\mu, pt\mu$ and $dt\mu$. As is well known there are 
four bound $S(L = 0)-$states in these three ions: three ground $S(L = 
0)-$states (one in each of these ions) and one excited $S(L = 0)-$state in the 
heavy $dt\mu$ ion. In this study we want to investigate the hyperfine structure 
and determine the hyperfine structure splittings for each of these bound states 
by using highly accurate expectation values of the delta-functions obtained in 
recent highly accurate numerical computations. 

The general formula for the hyperfine structure splitting $(\Delta H)_{h.s.}$ (or 
hyperfine splitting, for short) for an arbitrary three-body system is written as 
the sum of the three following terms. Each of these terms is proportional to the 
product of the factor $\frac{2 \pi}{3} \alpha^2$ and expectation value of the 
corresponding (interparticle) delta-funtion. The third (additional) factor 
contains the corresponding $g-$factors (or hyromagnetic ratios) and scalar 
product of the two spin vectors. For instance, for the $pd\mu$ ion this formula 
takes the form (in atomic units) (see, e.g., \cite{LLQ}, \cite{Fro02})
\begin{eqnarray}
 (\Delta H)_{h.s.} = \frac{2 \pi}{3} \alpha^2 \frac{g_p g_d}{m^2_p}
 \langle \delta({\bf r}_{pd}) \rangle ({\bf s}_p \cdot {\bf s}_d)+
 \frac{2 \pi}{3} \alpha^2 \frac{g_p g_{\mu}}{m_p m_{\mu}}
  \langle \delta({\bf r}_{p\mu}) \rangle ({\bf s}_p \cdot {\bf s}_{\mu}) 
 \nonumber \\
 + \frac{2 \pi}{3} \alpha^2 \frac{g_d g_{\mu}}{m_p m_{\mu}}
  \langle \delta({\bf r}_{d\mu}) \rangle ({\bf s}_d \cdot {\bf s}_{\mu})
 \label{e1}
\end{eqnarray}
where $\alpha = \frac{e^2}{\hbar c}$ is the fine structure constant, $m_{\mu}$ 
and $m_p$ are the muon and proton masses, respectively. The factors $g_{\mu}, 
g_{p}$ and $g_{d}$ are the corresponding $g-$factors. The expression for
$(\Delta H)_{h.s.}$ is, in fact, an operator in the total spin space which has 
the dimension $(2 s_p + 1) (2 s_d + 1) (2 s_{\mu} + 1) = 12$. In our 
calculations we have used the following numerical values for the constants and 
factors in Eq.(\ref{e1}): $\alpha = 7.297352586 \cdot 10^{-3}, m_p = 
1836.152701 m_e, m_{\mu} = 206.768262 m_e$ and $g_{\mu} = -2.0023218396$. The 
$g-$factors for the proton and deuteron are deteremined from the formulas: $g_p 
= \frac{{\cal M}_d}{I_p}$ and $g_d = \frac{{\cal M}_d}{I_d}$, where ${\cal M}_p 
= 2.792847386$ and ${\cal M}_d = 0.857438230$ are the magnetic moments (in 
nuclear magnetons) of the proton and deuteron, respectively. The spin of the 
proton and deuteron is designated in Eq.(\ref{e1}) as $I_p = \frac12$ and $I_d 
= 1$.  

The analogous formula for the hyperfine structure splitting in the $pt\mu$ ion 
takes the form 
\begin{eqnarray}
 (\Delta H)_{h.s.} = \frac{2 \pi}{3} \alpha^2 \frac{g_p g_t}{m^2_p}
 \langle \delta({\bf r}_{pt}) \rangle ({\bf s}_p \cdot {\bf s}_t) +
 \frac{2 \pi}{3} \alpha^2 \frac{g_p g_{\mu}}{m_p m_{\mu}}
  \langle \delta({\bf r}_{p\mu}) \rangle ({\bf s}_p \cdot {\bf s}_{\mu}) 
 \nonumber \\
 + \frac{2 \pi}{3} \alpha^2 \frac{g_t g_{\mu}}{m_p m_{\mu}}
  \langle \delta({\bf r}_{t\mu}) \rangle ({\bf s}_t \cdot {\bf s}_{\mu})
 \label{e3}
\end{eqnarray}
where $g_t = \frac{{\cal M}_t}{I_t}$, where ${\cal M}_t = 2.9789624775$ is the
magnetic moment of the triton expressed in the nuclear magnetons and $I_t = 
\frac12$ is the spin of the triton (or tritium nucleus). The formula for the 
hyperfine structure splitting in the $dt\mu$ ion is
\begin{eqnarray}
 (\Delta H)_{h.s.} = \frac{2 \pi}{3} \alpha^2 \frac{g_d g_t}{m^2_p}
 \langle \delta({\bf r}_{dt}) \rangle ({\bf s}_d \cdot {\bf s}_t) +
 \frac{2 \pi}{3} \alpha^2 \frac{g_d g_{\mu}}{m_p m_{\mu}}
  \langle \delta({\bf r}_{d\mu}) \rangle ({\bf s}_d \cdot {\bf s}_{\mu}) 
 \nonumber \\
 + \frac{2 \pi}{3} \alpha^2 \frac{g_t g_{\mu}}{m_p m_{\mu}}
  \langle \delta({\bf r}_{t\mu}) \rangle ({\bf s}_t \cdot {\bf s}_{\mu})
 \label{e3}
\end{eqnarray}
where all values are defined above. The same formula can be applied to determine 
the hyperfine structure spllitting in the excited $S(L = 0)-$state of the $dt\mu$ 
ion. The only difference in the hyperfine structure splittings determined for the 
ground and excited states of the $dt\mu$ ion can be related with the expectation 
values of interparticle delta-functions. 

In our computations of the muonic molecular ions performed recently \cite{FrWa2011}
we have determined the expectation values of all delta-functions which are needed 
in Eqs.(\ref{e1}) - (\ref{e3}). The corresponding expectation values are shown in 
Table I. These values have been determined in muon atomic units where $m_{\mu} = 1, 
\hbar = 1, e = 1$. They must be re-calculated to the regular atomic units ($m_e = 
1, \hbar = 1, e = 1$) which are used in the formulas, Eqs.(\ref{e1}) - (\ref{e3}),
to determine the hyperfine structure splittings. In these calculations we have used 
the trial wave functions with N = 3300, 3500, 3700 and 3840 exponential basis 
functions (for more details, see \cite{FrWa2011}). The expectation values of all
interparticle delta-functions computed for the ground $S(L = 0)-$state of the 
$pd\mu$ ion are shown in Table I. The overall convergence rates of the 
delta-functions computed for each bound state in the $pt\mu$ and $dt\mu$ ions are 
very similar to the results shown in Table I.   

These expectation values of the $\delta({\bf r}_{ij})$-functions were used in the
formulas Eqs.(\ref{e1}) - (\ref{e3}) to determine the hyperfine structure splittings 
of the bound $S(L = 0)-$states of the $pd\mu, pt\mu$ and $dt\mu$ ions. Numerical 
values of the corresponding hyperfine structure splittings can be found in Tables II
and III. Note that these values are usually given in $MHz$, while the values of 
$(\Delta H)_{h.s.}$  which follow from Eqs.(\ref{e1}) - (\ref{e3}) are expressed in 
atomic units. To re-calculate them from atomic units to $MHz$ the conversion factor 
6.57968392061 $\cdot 10^9$ $MHz/a.u.$ was used \cite{CRC}.

In general, the $pd\mu$ and $dt\mu$ ions have similar hyperfine structure. In
particular, in each of these ions one finds twelve spin states which are separated 
in the four following groups: (1) the group with $J = 2$ (five states), (2) the 
group with $J = 1$ (three states), (3) the group of one state with $J = 0$ (one 
state) and (4) the group with $J = 1$ (three states). Here and everywhere below 
the notation $J$ stands for the total spin (or total momentum, for the $S(L = 
0)-$states) of the three-body ion. The states with $J = 2$ have the maximal energy, 
while the energy of the states from the fourth group is minimal. The corresponding 
splittings $\Delta_{12}, \Delta_{23}$ and $\Delta_{34}$ can be found in Table II 
for each bound state in the $pd\mu$ and $dt\mu$ ions.

The hyperfine structure of the ground state in the $pt\mu$ ion is completely 
different (see Table III) from the hyperfine structures of the $pd\mu$ and $dt\mu$ 
ions discussed above. It follows from the fact that the spin of the triton equals 
$\frac12$, while the spin of the deuteron (or deuterium nucleus) equals 1. In the 
case of the ground state in the $pt\mu$ ion one finds only eight spin states which 
are separated into three different groups: (1) the group of four states with $J = 
\frac32$, (2) the group of two states with $J = \frac12$ and (3) the group of two 
states with $J = \frac12$. The group (1) has the maximal energy, while the energy of 
the states from the third group is minimal. The corresponding values of the hyperfine 
structure splittings in the ground state of the $pt\mu$ ion are $\Delta_{12}$ = 
1.3400149$\cdot 10^7$ $MHz$ and $\Delta_{23}$ = 3.3518984$\cdot 10^7$ $MHz$.

Thus, we have determined the hyperfine structure splitting in the bound $S(L = 
0)-$states of the $pd\mu, pt\mu$ and $dt\mu$ ions. The hyperfine structure splitting
of the first excited $S(L = 0)-$state in the $dt\mu$ ion is considered too. This 
excited state is traditionally designated by an additional asterisk, i.e. $(dt\mu)^*$. 
In our calculations we used the highly accurate expectation values of all 
interparticle delta-functions obtained in recent computations \cite{FrWa2011}. In 
general, it is very interesting to compare the numerical values of the hyperfine 
structure splittings $\Delta_{12}, \Delta_{23}$ and $\Delta_{34}$ determined for the 
different muonic ions (see Table II).

\newpage 

\begin{table}[tbp]
    \caption{The convergence of the $\langle \delta_{32} \rangle, 
             \langle \delta_{31} \rangle$ and $\langle \delta_{31} \rangle$
             expectation values for the ground (bound) $S(L = 0)-$state of 
             the $pd\mu$ molecular ion (in muon-atomic units).}
      \begin{center}
      \begin{tabular}{llll}
        \hline\hline
 $N$ & $\langle \delta_{32} \rangle$ & $\langle \delta_{31} \rangle$ & $\langle \delta_{21} 
 \rangle$ \\
      \hline
  3300 & 1.73456203087$\cdot 10^{-1}$ & 1.17709732798$\cdot 10^{-1}$ & 1.46169407$\cdot 10^{-5}$ \\

  3500 & 1.73456202965$\cdot 10^{-1}$ & 1.17709733128$\cdot 10^{-1}$ & 1.46169370$\cdot 10^{-5}$ \\

  3700 & 1.73456202754$\cdot 10^{-1}$ & 1.17709733014$\cdot 10^{-1}$ & 1.46169377$\cdot 10^{-5}$ \\

  3840 & 1.73456202768$\cdot 10^{-1}$ & 1.17709733160$\cdot 10^{-1}$ & 1.46169383$\cdot 10^{-5}$ \\
       \hline\hline
   \end{tabular}
   \end{center}
   \end{table}
\begin{table}[tbp]
    \caption{The levels of hyperfine structure $\epsilon$ and hyperfine structure splittings 
             $\Delta$ in the bound $S(L = 0)-$states of the $pd\mu$ and $dt\mu$ ions 
             (in $MHz$).}
      \begin{center}
      \begin{tabular}{lllll}
        \hline\hline
$\epsilon_{J=2}(pd\mu)$ &  1.2519350851$\cdot 10^7$ & ----- & ------------ \\ 

$\epsilon_{J=1}(pd\mu)$ &  9.3058194294$\cdot 10^6$ & $\Delta(J=2 \rightarrow J=1)$ & 3.2135314217$\cdot 10^6$ \\ 

$\epsilon_{J=0}(pd\mu)$ & -2.1222395094$\cdot 10^7$ & $\Delta(J=1 \rightarrow J=0)$ & 3.0528214524$\cdot 10^7$ \\

$\epsilon_{J=1}(pd\mu)$ & -2.3097272483$\cdot 10^7$ & $\Delta(J=0 \rightarrow J=1)$ & 1.8748773889$\cdot 10^6$ \\
      \hline
      \hline
$\epsilon_{J=2}(dt\mu)$ &  1.8919590437$\cdot 10^7$ & ----- & ------------ \\ 

$\epsilon_{J=1}(dt\mu)$ &  1.5985479092$\cdot 10^6$ & $\Delta(J=2 \rightarrow J=1)$ & 2.9341113453$\cdot 10^6$ \\ 

$\epsilon_{J=0}(dt\mu)$ & -3.4439378258$\cdot 10^7$ & $\Delta(J=1 \rightarrow J=0)$ & 5.0424857350$\cdot 10^7$ \\

$\epsilon_{J=1}(dt\mu)$ & -3.6038337067$\cdot 10^7$ & $\Delta(J=0 \rightarrow J=1)$ & 1.5989588085$\cdot 10^6$ \\
      \hline
      \hline
$\epsilon_{J=2}(dt\mu)^{*}$ &  1.8609555434$\cdot 10^7$ & ----- & ------------ \\ 

$\epsilon_{J=1}(dt\mu)^{*}$ &  1.6554331952$\cdot 10^6$ & $\Delta(J=2 \rightarrow J=1)$ & 2.0552234821$\cdot 10^6$ \\ 

$\epsilon_{J=0}(dt\mu)^{*}$ & -3.4859818025$\cdot 10^7$ & $\Delta(J=1 \rightarrow J=0)$ & 5.1414149977$\cdot 10^7$ \\

$\epsilon_{J=1}(dt\mu)^{*}$ & -3.5950318334$\cdot 10^7$ & $\Delta(J=0 \rightarrow J=1)$ & 1.0905003094$\cdot 10^6$ \\
      \hline\hline
   \end{tabular}
   \end{center}
   \end{table}
\begin{table}[tbp]
    \caption{The levels of hyperfine structure $\epsilon$ and hyperfine structure splittings
             $\Delta$ in the ground $S(L = 0)-$state of the $pt\mu$ ion (in $MHz$).}
      \begin{center}
      \begin{tabular}{lllll}
        \hline\hline
$\epsilon_{J=\frac32}(pt\mu)$ &  1.5079820356$\cdot 10^7$ & ----- & ------------ \\ 

$\epsilon_{J=\frac12}(pt\mu)$ &  1.6796717260$\cdot 10^6$ & $\Delta(J=\frac32 \rightarrow J=\frac12)$ & 1.3400148630$\cdot 10^7$ \\ 

$\epsilon_{J=\frac12}(pt\mu)$ & -3.1839312439$\cdot 10^7$ & $\Delta(J=\frac12 \rightarrow J=\frac12)$ & 3.3518984165$\cdot 10^7$ \\
      \hline
      \hline
   \end{tabular}
   \end{center}
   \end{table}

\end{document}